**Structural gradients and strain partitioning across the mouse Achilles tendon enthesis revealed by *in situ* X-ray scattering**

Isabella Silva Barreto[1], Moritz L. Stammer[1], Moritz P.K. Frewein[1], Claire Camy[2,3], Juraj Todt[4,5], Michael Meindlhumer[4], Jozef Keckes[4], Stefano Checchia[6], Sandrine Roffino[2], Martine Pithioux[2,3,7,†], Tilman A. Grünewald[1,†,*]

**Affiliations:**

[1] Aix-Marseille Université, CNRS, Centrale Med, Institut Fresnel, Marseille, France

[2] Aix-Marseille Université, CNRS, ISM, Institut des Sciences du Movement, Marseille, France

[3] Aix-Marseille Université, APHM, CNRS, ISM, Mecabio Facility, Anatomy Laboratory, Timone, Marseille, France

[4] Department of Materials Science, Montanuniversität Leoben, Franz Josef Straße 18, 8700 Leoben, Austria

[5] Erich Schmid Institute for Materials science, Austrian Academy of the Sciences, Jahnstraße 12, 8700 Leoben, Austria

[6] European Synchrotron Radiation Facility, Grenoble, France

[7] Aix-Marseille Université, APHM, CNRS, ISM, Sainte-Marguerite Hospital, Institute for Locomotion, Marseille, France

[†]These authors contributed equally

*To whom correspondence should be addressed:

Tilman A. Grünewald, Email: tilman.grunewald@fresnel.fr

Institut Fresnel

52 Avenue Escadrille Normandie Niemen, 13013 Marseille, France

**ABSTRACT**

The enthesis is the insertion site of tendon into bone and exhibits a high mechanical durability despite the large mismatch in material properties between the two tissues. This durability stems from gradients in composition, structure and organization on multiple hierarchical length scales. Despite extensive research on enthesis structure and mechanics, the local deformation mechanisms are poorly understood. Synchrotron scanning small- and wide-angle X-ray scattering was combined with *in situ* tensile testing of the mouse Achilles tendon enthesis to extensively map the mechanical response of the collagen fibrils and molecules as well as the hydroxyapatite mineral particles and crystals. Gradients in nano- and molecular scale structure and a stronger and more immediate deformation response towards the interface compared to further away were observed in both the soft and mineralized tissue. The strain decreased progressively across hierarchical levels; with an applied tissue strain of 20% the nanoscale fibrils were strained by ~1-2%, the collagen molecules by ~0.5% and the hydroxyapatite crystals by ~0.05%, thus following an approximate ratio of 1 : 0.1 : 0.01 : 0.001. These results show that load transfer across the enthesis is both spatially heterogeneous and hierarchy-dependent. This indicates that the graded attachment accommodates deformation through region-specific load sharing and hierarchical strain partitioning, consistent with a contribution from dissipation within the non-collagenous matrix. In doing so, the enthesis can mitigate stress concentrations and maintain mechanical integrity across the tendon-to-bone transition.

## 1 INTRODUCTION

The enthesis, the insertion site of a tendon or ligament into bone, enables load transfer between the two mechanically dissimilar materials: compliant tendons and stiff bone[1,2]. Such a large mismatch in material properties over a very short distance (~200 µm in the mouse Achilles tendon enthesis[3,4]) would normally be expected to generate stress concentrations[5] and promote failure. Instead, the enthesis is remarkably durable and less prone to failure than either tendon or bone themselves[6]. This is generally attributed to gradual changes in composition, organization and structure across the interface over multiple hierarchical length scales[5–10]. When injured however, the enthesis regenerates poorly and surgical repair often fails[2,7,11]. A better mechanistic understanding of how this interface accommodates load is therefore needed. Despite extensive research on the enthesis structure and mechanics, the local deformation mechanisms that govern load transfer across regions and length scales remain insufficiently understood.

Fibrocartilaginous entheses, such as the Achilles tendon insertion, connect tendon to bone through a transition region of uncalcified- (UFc) and mineralized fibrocartilage (MFc)[2,12] (Fig 1.A). Across this transition, the tissues exhibit marked differences in composition, structure and viscoelastic mechanical behavior[13–16]. Tendon and bone are both rich in collagen type I, whereas fibrocartilaginous tissue contains predominantly collagen type II. Tendon and fibrocartilage are highly hydrated, containing ~70% water, whereas bone only contains about 20%. In addition, tendons include non-collagenous proteins such as elastin and both tendons and fibrocartilage contain a large portion of proteoglycans, which both contribute substantially to their mechanical response[17]. In mineralized fibrocartilage and bone, the tissue is further reinforced by a mineral phase consisting primarily of carbonated hydroxyapatite (HA). At the nano- and molecular scale however, these tissues share a common hierarchical organization (Fig 1.B). Collagen molecules consist of different amino-acid chains assembled into triple helices which are cross-linked and organized into fibrils in a quarter staggered periodic arrangement, which in turn are further organized into microscale fibers. Within the mineralized tissues, HA mineral particles are interspersed within and across the collagen fibrils[18–21], with the mineral distribution changing from predominantly extrafibrillar near the tidemark, to mixed extra- and intrafibrillar mineral closer to the bone[22].

Because collagen fibrils, collagen molecules, HA particles and HA crystals all possess pronounced periodic order, these structural levels can be probed efficiently using X-ray scattering[23]. Small-angle X-ray scattering (SAXS) provides information about nanoscale arrangements such as the collagen fibril axial d-spacing (~64-67 nm) and lateral packing distance (~1-1.5 nm)[13,24–26], as well as the size of the mineral particles (~2-4 nm)[27,28]. Wide-angle X-ray scattering (WAXS) in turn provides information on the distance between amino acids in the collagen molecules (one full helical rotation, ~2-3 Å)[24–26,29], as well as the crystalline properties of the mineral components such as lattice spacing and crystallite size[30,31]. X-ray diffraction has been extensively used to study the nanoscale structure and mechanics collagen and HA mineral in tendons and bone[13,32–42]. Although studies of collagen molecular mechanics have been more limited and focused primarily on skin[13,24,25], X-ray scattering experiments have been central to build our current understanding of collagen and mineral deformation mechanisms. In tendon, fibrils sustain strains that are substantially lower than the applied tissue strains[40–47] and far below failure strains of individual fibrils[48–50]. This discrepancy points towards strain partitioning across hierarchical levels, likely mediated in part by shear forces and sliding within the interfibrillar and interfiber matrix[51–55]. Most such studies however, have relied on single spot measurements[39–42,45,46,56–58] which allows high time resolution within a given radiation damage budget, but sacrifices spatial information. More recently, 2D scanning SAXS approaches have begun to resolve the spatially heterogeneous nanoscale collagen mechanics in the cartilage-bone unit of the knee[59], in healing Achilles tendons[60] and in keloid scars[61]. By contrast, SAXS/WAXS of the enthesis remain scarce. At larger length scales, X-ray micro-computed tomography (μCT) has shown increased[1,62] microscale strains at the insertion site compared to the tendon, but that there are localized regions where it is instead decreased[1]. Recent work on the mouse Achilles enthesis further showed substantial insertion-specific compressive and impingement-related strains during dorsiflexion[63,64]. The strain response depends mostly on the angle of the applied load[1] and fibers bending close to the mineralized tissue at low loads[62,65]. More broadly, recent work has also shown that the mouse Achilles enthesis adapts structurally and mechanically to increased loading, underscoring the load sensitivity of this graded attachment[66,67]. Together, these studies underline that the enthesis exhibits a highly complex and spatially heterogenous mechanical response that remains debated, especially at the nano- and microscale.

Here, we combine 2D scanning SAXS and WAXS with *in situ* tensile loading of mouse Achilles tendon enthesis using a custom-built loading device (Fig 2.A). By mapping the mechanical response of the collagen fibrils and molecules together with HA mineral particles and crystals across a large field of view, we resolved how local deformation is partitioned across regions and hierarchical length scales. This approach revealed a spatially heterogeneous multiscale mechanical response across tendon, interface and bone, which provides direct insight into how the enthesis accommodates load across this graded biological junction.

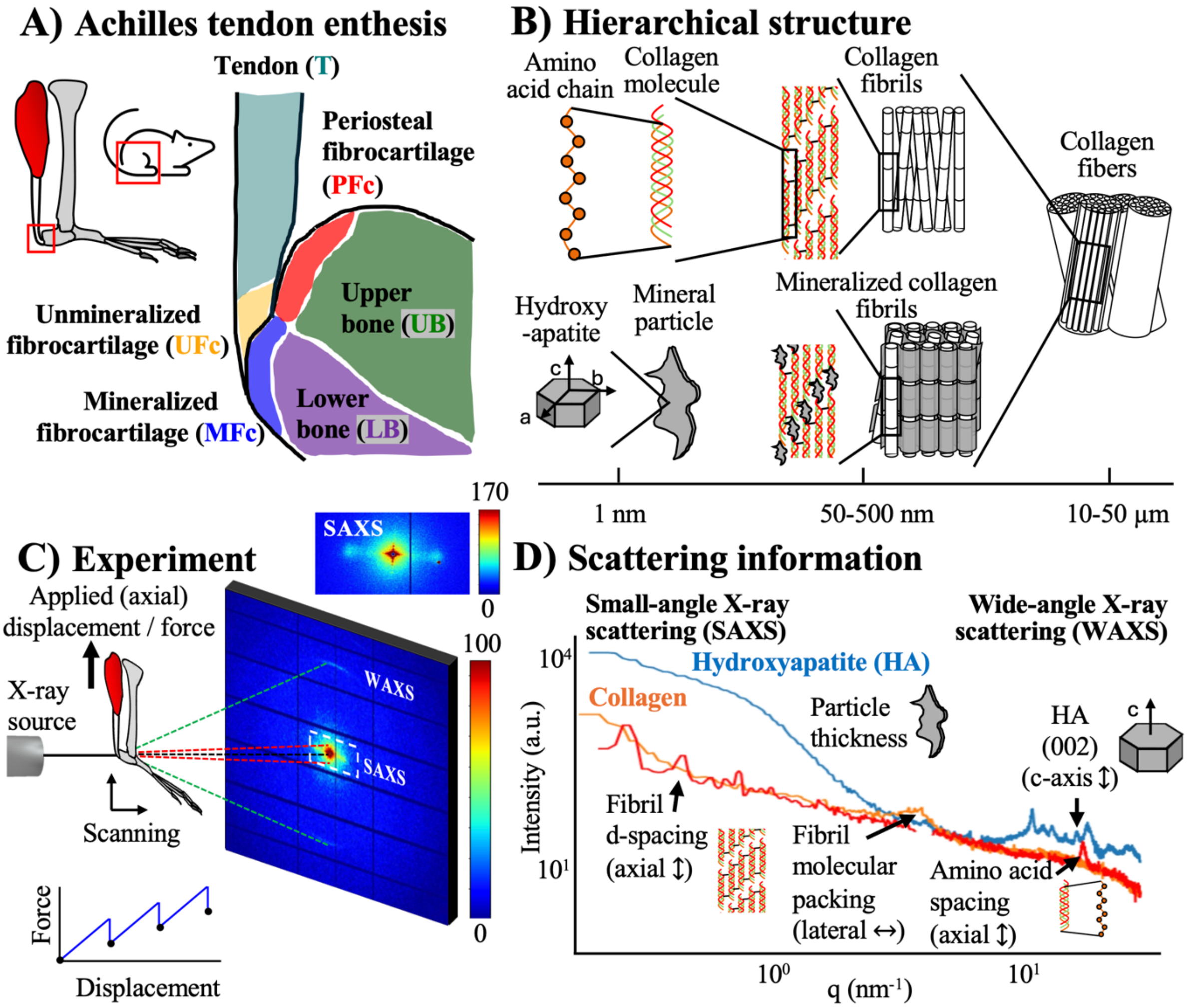

**Figure 1. Overview of enthesis structure and methods**. A) Illustration of a fibrocartilaginous enthesis such as the mouse Achilles tendon enthesis, indicating the Achilles tendon (T), unmineralized- (UFc), mineralized- (MFc) and periosteal fibrocartilage (PFc), as well as the upper (UB) and lower (LB) part of the calcaneal bone. B) Schematic of the hierarchical structure of the collagen and mineral tissues in the enthesis. C) Schematic of the 2D scanning SAXS/WAXS experiment at the synchrotron, showing a representative collagen scattering pattern with a zoom-in of the small-angle region. D) Representative radially integrated intensity of the collagen signal in the axial (red) and lateral (orange) directions as well as that of the HA mineral. The investigated parameters of the mineral (SAXS shoulder related to particle thickness and WAXS (002) reflection related to the crystal lattice spacing along the c-axis) and collagen (SAXS 5$^{th}$ collagen reflection related to fibril axial d-spacing and WAXS reflections related to fibril lateral molecular packing and axial amino acid spacing) tissues are indicated.

## 2 RESULTS

### 2.1 Macroscale mechanics

To investigate the mechanical response of the enthesis, mouse enthesis samples were mounted in a custom-made tensile device (Fig 2.A) on the ID15A beamline at the European Synchrotron Radiation Facility (ESRF), France. The samples were preloaded to 2 N and tested in axial tension using three incremental displacement steps of 200 µm each. Each step was followed by either 1 min (N=5) or 15 min (N=5) relaxation to assess the effect of viscoelastic relaxation. Each step resulted in approximately 20% tissue strain (Table 1, tissue scale, Fig 2.B). In both groups, the peak force increased with increased displacement, while stiffness only showed a slight increase in the 1min relaxation group (Fig 2.C). No significant effect of displacement was observed on the relaxation ratio or fast relaxation time. No significant differences in these parameters were observed between the two relaxation times, except for a shorter relaxation time at the second displacement step after 1 min compared to 15 min (Fig 2.C). As the variability across samples in the force relaxation response overlapped between both relaxation times, the samples from both were pooled for the structural analysis of the smaller length scales.

**Table 1. Strain distribution across hierarchical length scales.** Comparison of average strain at each step at the tissue scale for 1 and 15min relaxation, as well as within the region closest to the interface (Uncalcified fibrocartilage UFc, mineralized fibrocartilage MFc) compared to further away (Achilles tendon T, lower bone LB) at the nanoscale and molecular scale for all samples. The average ratio of the nano- or molecular scale strain compared to the corresponding applied tissue strain (within each sample).

| | **Tissue scale** | | **Nanoscale** | | | | **Molecular scale** | | **Crystal scale** | |
|---|---|---|---|---|---|---|---|---|---|---|
| | Applied strain (%) | | Fibril (axial) strain (%) | | Fibril (lateral) strain (%) | | Amino acid (axial) strain (%) | | HA $d_{002}$ strain (%) | |
| Step | **1min** | **15min** | **UFc** | **T** | **UFc** | **T** | **UFc** | **T** | **MFc** | **LB** |
| 1 | 20 ± 5 | 17 ± 5 | 0.8 ± 0.4 | 0.4 ± 0.3 | -2.2 ± 0.8 | -1.7± 0.6 | 0.6 ± 0.4 | 0.6 ± 0.3 | 0.08 ± 0.02 | 0.04 ± 0.01 |
| **ratio** | **1** | | **0.04** | **0.02** | **0.12** | **0.09** | **0.03** | **0.04** | **0.004** | **0.002** |
| 2 | 41 ± 10 | 36 ± 2 | 1.2 ± 0.5 | 0.9 ± 0.5 | -2.7 ± 0.5 | -2.1 ± 0.6 | 0.6 ± 0.6 | 0.9 ± 0.6 | 0.11 ± 0.03 | 0.06 ± 0.02 |
| **ratio** | **1** | | **0.03** | **0.02** | **0.07** | **0.05** | **0.02** | **0.02** | **0.003** | **0.002** |
| 3 | 62 ± 21 | 53 ± 0.0 | 1.0 ± 0.5 | 1.1 ± 0.7 | -2.5 ± 0.2 | -2.2 ± 0.2 | 0.2 ± 0.3 | 1.5 ± 0.02 | 0.10 ± 0.02 | 0.06 ± 0.01 |
| **ratio** | **1** | | **0.02** | **0.02** | **0.04** | **0.04** | **0.003** | **0.03** | **0.002** | **0.001** |

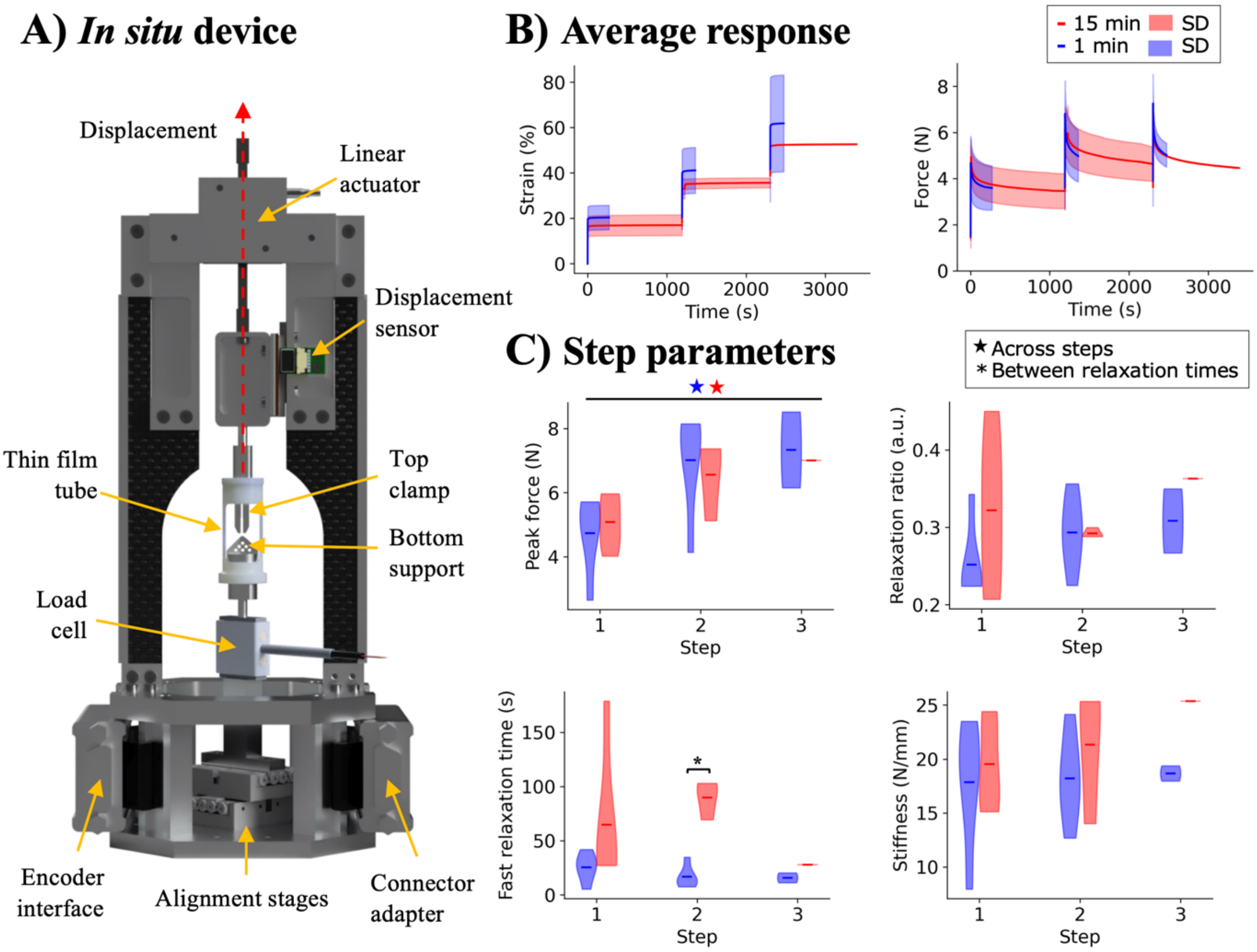

**Figure 2. Macroscale mechanical response.** A) Design of the custom-made *in situ* loading device. B) Average force relaxation response of 1 (blue) vs 15 (red) minutes relaxation, indicating the group mean (solid line) and standard deviation across samples (shaded area). C) Average parameters at each strain step, indicating the group mean (solid line) as well as the distribution of data (the width of the shaded area represents the proportion of data at that value). Statistical significance (i.e. probability value $p < 0.05$) is indicated across steps within a relaxation time (color-coded star) and between relaxation times at each step (asterisk).

### 2.2 Unmineralized tissue mechanics

To characterize the multiscale mechanical response of the tissue components, 2D projections (1.5 x 1 mm, 20 µm steps, 4 ms exposure time) covering the Achilles tendon, enthesis and posterior parts of the calcaneal bone were acquired at preload and after relaxation at each displacement step using scanning SAXS/WAXS (Fig 1.C). One sample was excluded from the results due to beam alignment issues for the SAXS/WAXS analysis, but was retained for the mechanics analysis. From the SAXS signal (Fig 1.D), the nanoscale mechanical response of the collagen fibril was characterized through the axial d-spacing ($d_{fibril-A}$) and lateral molecular packing ($d_{fibril-L}$), their heterogeneity and detected amount / intrafibrillar order, their orientation and degree of orientation. From the WAXS signal (Fig 1.D), the molecular-scale mechanical response of the collagen was characterized through the axial amino acid spacing ($d_{amino}$), its heterogeneity, detected amount, orientation and degree of orientation.

### 2.2.1 Nanoscale collagen fibrils

At the nanoscale, structural differences were observed between regions (Fig 3.A). At preload, collagen fibrils in the unmineralized tissues exhibited a slightly larger axial d-spacing in the distal region closer to the mineral interface than in the main part of the tendon. The average lateral packing distance was similar in both regions (Fig 3.A ii) except for within the 50µm closest to the interface (Fig 3.A iii, preload). Additionally, both parameters exhibited a slight gradient going from the tendon towards the mineral interface, with the axial d-spacing increasing and the lateral packing distance decreasing (Fig 3.A iii, preload). While the heterogeneity of the axial d-spacing was slightly increased in the closer to the mineral interface compared to the main part of the tendon, no differences were observed in the lateral packing (SI Fig S2.A). Additionally, the main part of the tendon exhibited a slightly intrafibrillar order in both axial and lateral directions, as well as a significantly higher degree of orientation (SI Fig S2.A).

With increasing displacement steps, collagen fibrils exhibited approximately 1-2% strain, with tensile deformation in the axial direction (along the applied load) and lateral contraction (Table 1, nanoscale, SI Fig S1.A). In both regions, the lateral contraction of the molecular packing distance within the fibril exceeded the axial extension, resulting in a fibril contraction ratio larger than 1 (Table 2). Initially, the region closer to the mineral interface showed a stronger and more immediate response in both fibril axial d-spacing and the lateral molecular packing than the main part of the tendon (Fig 3.A ii-iii), which instead responded more incrementally. For the other axial fibril parameters, both regions showed similar trends with increased displacement, including a significant increase in strain heterogeneity and significant decrease in intrafibrillar order (SI Fig S2.A). For the other lateral fibril parameters, an effect of displacement steps was only observed in the intrafibrillar order of the fibrils in the main part of the tendon (SI Fig S2.A).

**Table 2. Fibril contraction ratio.** Comparison of average fibril contraction ratios of each step within the region closest to the interface (UFc) compared to the main part of the Achilles tendon (T).

| Region | Step 1 | Step 2 | Step 3 |
|---|---|---|---|
| UFc | 3.2 ± 2.3 | 3.1 ± 3.0 | 2.4 ± 1.4 |
| T | 5.7 ± 4.9 | 3.8 ± 2.9 | 3.6 ± 1.8 |

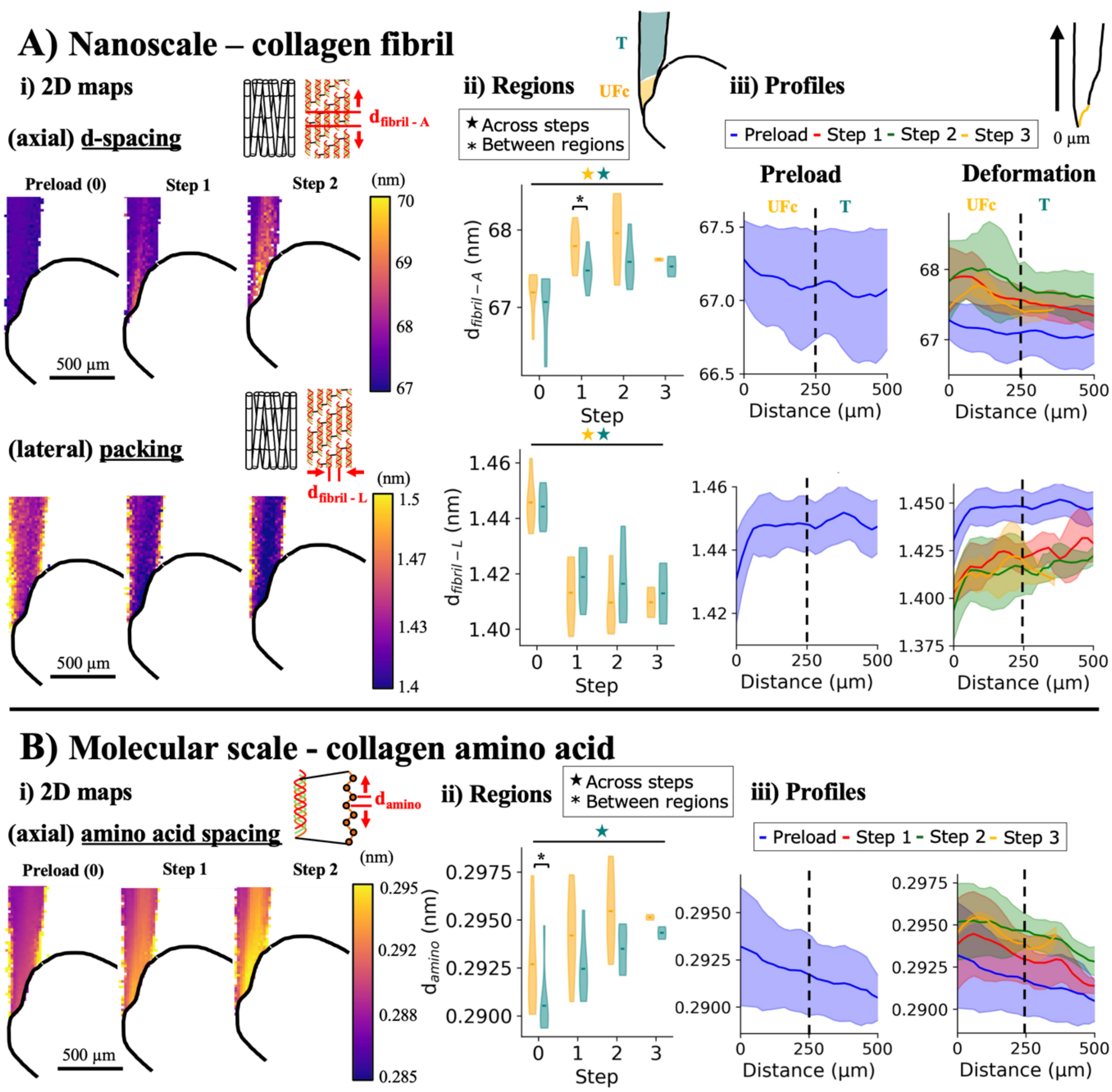

**Figure 3. Unmineralized tissue hierarchical results.** A) Nanoscale collagen fibril axial and lateral mechanical results and B) molecular scale collagen amino acid axial mechanical results with (i) representative 2D maps of the spacing in each point, (ii) average spacing within each region across displacement steps (0 = preload), showing the group mean (solid line) as well as the distribution of data (the width of the shaded area represents the proportion of data at that value), and (iii) profiles in spacing at preload as well as at each step as a function of distance to the mineral interface, showing the group mean (solid line) and standard deviation across samples (shaded area). Statistical significance ($p < 0.05$) is indicated across steps within a region (color-coded stars) and between regions at each step (asterisk).

### 2.2.2 Collagen molecules

On the molecular scale, structural differences were likewise observed between regions in the amino acid spacing of the collagen molecules (Fig 3.B). At preload, collagen molecules in the unmineralized tissues exhibited a significantly larger amino acid spacing in the distal region closer to the mineral interface than in the main part of the tendon (Fig 3.B i-ii). Similar to the nanoscale fibrils, this increase was gradual from the tendon towards the mineral interface (Fig 3.B iii, preload). While a slight increase in heterogeneity and degree of orientation closer to the mineral interface was observed, there was no difference in internal order (SI Fig S2.B).

In general, the collagen molecules exhibited approximately 0.5% tensile strain with the applied tissue displacement (Table 1, molecular scale, SI Fig S1.A). Those in the region closer to the mineral interface showed less incremental strain response compared to those in the main part of the tendon (Fig 3.B, ii-iii, SI Fig S1.A). No clear trends were observed in the strain heterogeneity, internal order or degree of orientation with increasing displacement steps (SI Fig S2.B).

### 2.3 Mineralized tissue mechanics

From the SAXS signal (Figure 1.D), mineral particle size ($T$), orientation and degree of orientation were quantified. From the WAXS signal (Figure 1.D), the HA crystallites were characterized through the lattice spacing of the HA (002) reflection, ($d_{002}$), i.e. along the c-axis, its orientation and degree of orientation.

#### 2.3.1 Nanoscale mineral particles

At the nanoscale, structural differences were observed between regions. Both cartilage regions (MFc and PFc) exhibited significantly smaller mineral particles than bone regions (LB and UB) (Fig 4.A ii). Interestingly, the mineral particles in the MFc exhibited a higher degree of orientation compared to all other regions, as well as a different predominant particle orientation (Fig 4.A ii). The mineral particles remained largely unchanged with displacement steps (Figure 4.A), apart from a significant re-orientation in the MFc (Fig 4.A ii).

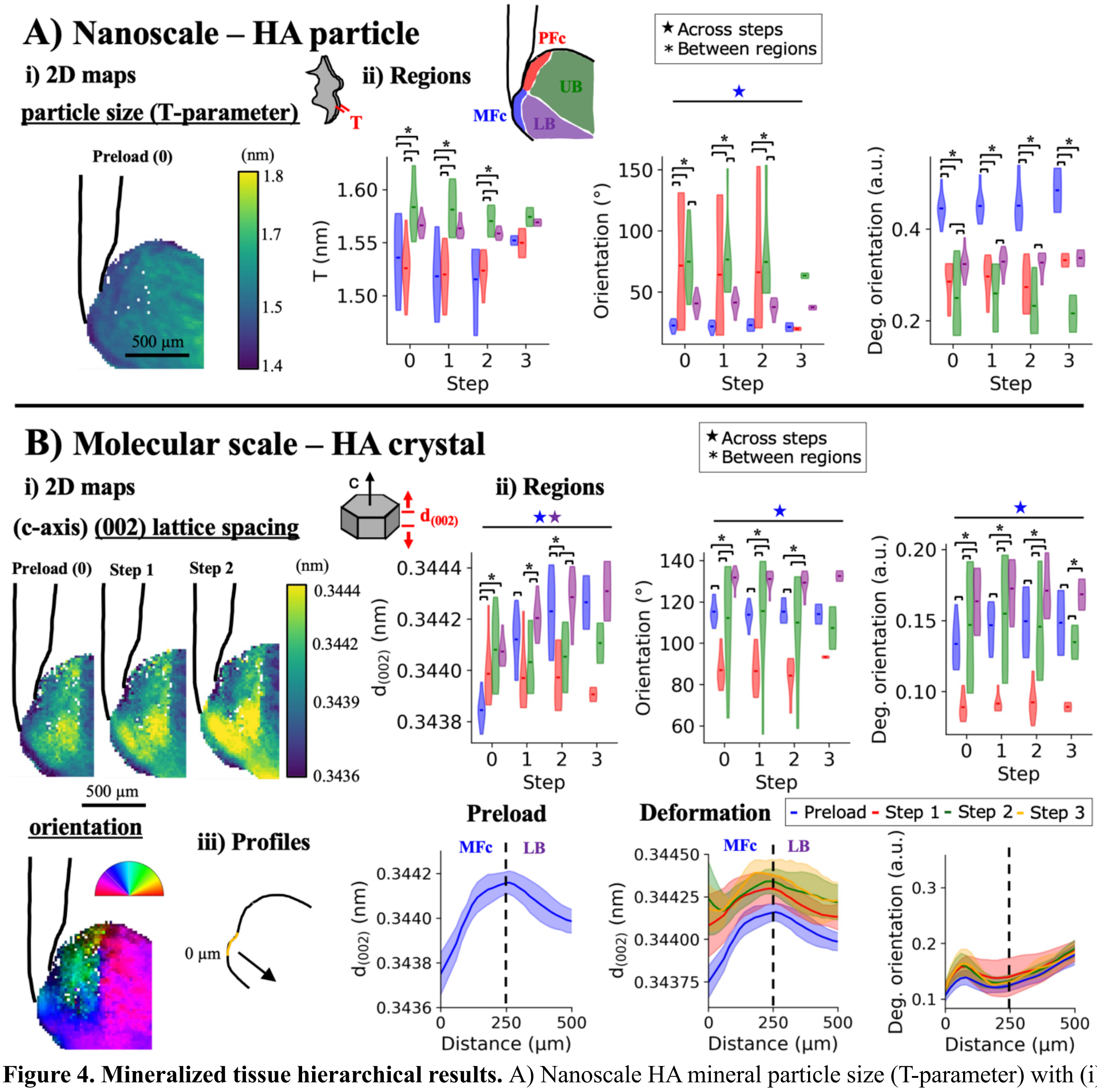

**Figure 4. Mineralized tissue hierarchical results.** A) Nanoscale HA mineral particle size (T-parameter) with (i) representative 2D map at preload and (ii) average size, orientation and degree of orientation within each region. Data is represented as mean (solid line) as well as distribution of data (the width of the shaded area represents the proportion of data at that value). B) HA crystal mechanical results, derived from HA $d_{002}$ with (i) representative 2D maps, (ii) average size, orientation and degree of orientation within each region across steps (0 = preload) and (iii) average size and degree of orientation profiles of each step as a function of distance to the soft tissue interface, showing the group mean (solid line) and standard deviation across samples (shaded area). Statistical significance ($p < 0.05$) is indicated across steps within a region (color-coded stars) and between regions at each step (asterisk).

### 2.3.2 HA crystals

At the crystal scale, structural differences between regions were observed in the HA $d_{002}$ spacing (Fig 4.B). At preload, the MFc exhibited a significantly smaller lattice spacing compared to all other regions (Fig 4.B ii). The lattice spacing gradually decreased towards the soft tissue interface (Fig 4.B iii). Additionally, the HA crystals in both the MFc and PFc exhibited a significantly different orientation and degree of orientation compared to the other regions (Fig 4.B i-ii).

In the MFc and LB, HA crystals exhibited approximately 0.05% c-axis related strain, estimated from changes in $d_{002}$, with increasing displacement (Table 1, crystal scale, SI Fig S1.A). The MFc started with a smaller spacing and showed a stronger strain response than the LB (Fig 4.B ii). At the larger displacement steps, both regions ultimately reached similar values. Additionally, the crystals of the MFc exhibited a significant increase in degree of orientation as well as reorientation with increasing displacement. The crystals of the PFc and UB remained unchanged with displacement.

## 3 DISCUSSION

This study provides the first in situ 2D full-field mapping of the nano- and molecular-scale deformation behavior across the mouse Achilles tendon enthesis, spanning tendon, cartilaginous interface and calcaneal bone. Generally, the enthesis is now recognized as a graded and mechanically specialized attachment whose structure and function are strongly shaped by mechanical load[9,10,12,14]. Earlier work further established the mature enthesis as a mechanically heterogeneous attachment with structurally distinct insertional zones [1,6,16,22,62]. This study extends this view by showing how these structurally different zones transfer mechanical load across hierarchical length scales. Specifically, we identify structural gradients already at preload, an earlier and stronger response of the interface-adjacent regions compared with tissues farther away, and a marked attenuation of strain from tissue to fibril, molecule, and crystal levels. These findings show that load transfer across the enthesis is both spatially heterogeneous and hierarchy-dependent. They further support a role for the non-collagenous matrix in modulating load transfer and provide a mechanistic framework for how this graded biological junction accommodates deformation.

### 3.1 Structural gradients and local response

The high failure resistance of the enthesis is widely attributed to its graded organization[1,6,14], which reduces stress concentrations during load transfer across the tendon-to-bone insertion[68]. In the mouse enthesis, previous work has already shown pronounced gradients in composition and mechanics, including a decrease in phosphate-to-collagen ratio and local changes in the elastic modulus from calcaneal bone toward the interface[4,16,22,68]. Additionally, micromechanical studies have suggested that the insertion region behaves differently from the surrounding tendon or bone[1,62–64]. The present results extend this enthesis-specific picture to smaller length scales. On the mineralized side, the graded organization is reflected by a decrease in HA c-axis lattice spacing from bone toward the unmineralized interface. This is consistent with previous reports of structural and functional heterogeneity in mineralized fibrocartilage near the Achilles tendon insertion[15,16]. Here, we also observe that this gradient is mirrored on the soft tissue side toward the mineral interface by an increase in collagen fibril d-spacing and amino-acid spacing, as well as a decrease in lateral fibril packing. The proposed interface-directed differences at preload are summarized schematically in Fig 5A. Interestingly, similar soft tissue gradients have previously been observed in the bone-cartilage unit[59,69,70]. These studies identified an increasing fibril d-spacing toward the mineral interface, which was interpreted as the presence of fibril pre-strain[69]. In the bone-cartilage unit, the gradient was linked to proteoglycan-rich regions and associated swelling effects[69]. The comparatively large fibrillar contraction ratio found in this study suggests that tensile loading is accompanied by substantial reorganization of the hydrated fibrillar environment. Because this ratio is derived from changes in scattering-based spacings, it should not be interpreted as a macroscopic Poisson ratio in the continuum mechanical sense. Nevertheless, such a fibrillar contraction response is consistent with the behavior of hydrated collagenous tissues[71–73], in which loading can drive fluid redistribution, matrix reorganization, and changes in fibril packing. The smaller fibrillar contraction ratio near the mineral interface may reflect additional constraints imposed by the graded attachment, including fiber splay, local mineral reinforcement, and proteoglycan-rich matrix components that modulate hydration and interfibrillar sliding. In general, the importance of several different types of proteoglycans for soft tissue biomechanics has been demonstrated[17]. Depletion of proteoglycans in tendons has led to both structural changes such as collagen cross-linking, diameter reduction and ossifications, as well as mechanical changes such as reduced tensile strength, stiffness and impaired viscoelasticity. On the nanoscale, depletion of proteoglycans and their glycosaminoglycan (GAG) side chains has been shown to increase collagen stretch and reduce fibril sliding at high strain levels[74], while also increasing stress relaxation[75].

Beyond suggesting a pre-strained fibrillar state similar to that described in the bone-cartilage unit, our data also point to an asymmetric spatial response of the enthesis. This suggests that the region close to the interface is already in a different structural state at preload and thus shows a larger response at the first load step compared to tendon and bone which has a more gradual response to load steps. Because the tendon fibers insert directly into bone[6], and mineral reinforcement decreases from bone through MFc towards the interface, the tissue closest to the insertion is expected to accommodate more deformation than the surrounding tendon or bone. In addition to this, the microscale collagen fibers have been shown to unravel into thinner, more splayed fibers at the interface[1], which most likely aids in further distributing the applied load. In the present 2D projection, a distinct orientation change was resolved primarily on the mineralized side of the interface, where MFc mineral particles exhibited significant reorientation with increasing displacement. The enthesis therefore most likely does not just act as a passive transition zone, but as a graded interface in which deformation is taken up earlier near the insertion and redistributed at larger load across the adjacent tissues. However, this apparent dependence on its gradients thus renders the whole system sensitive to alterations in the structure, composition or mechanical properties.

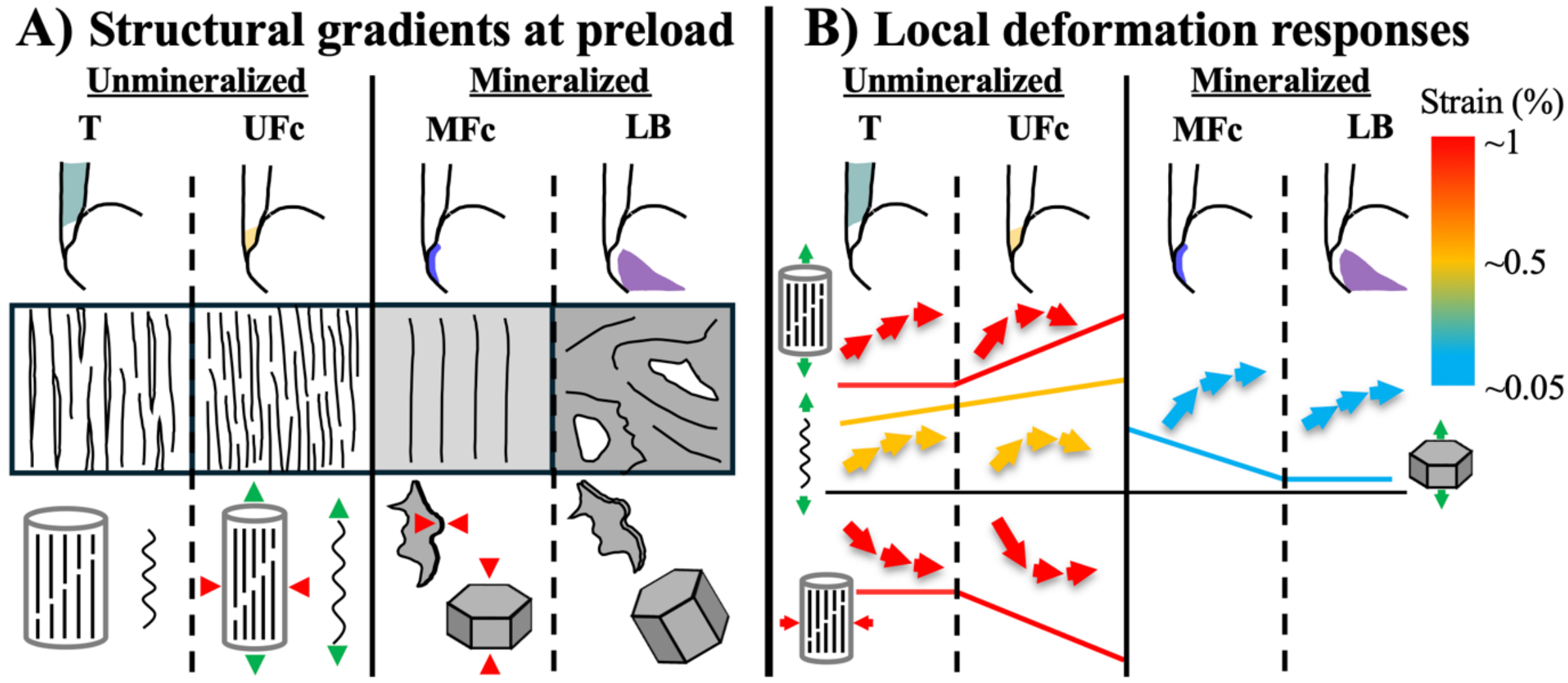

**Figure 5. Summary of local structure and mechanical behaviors.** A) Structural differences of the investigated nano-, molecular and crystal scale load bearing components (collagen and mineral) between the regions at 2N preload. B) Hierarchical distribution of strain between load bearing components as well as between regions. indicating the magnitude (color) and direction of their respective deformation upon loading (above midline = tensile, below = compression). The nature of their different deformation response within each region is highlighted with schematic arrows related to each displacement step.

### 3.2 Strain partitioning across hierarchical length scales

The strains measured in collagen fibrils, collagen molecules, and HA crystals were several orders of magnitude lower than the applied tissue strain and decreased progressively across hierarchical levels, following an approximate tissue : fibril : molecule : crystal ratio of 1 : 0.1 : 0.01 : 0.001. Thus, only a small fraction of the applied tissue deformation reached the load-bearing structures investigated here, indicating that most deformation was accommodated elsewhere in the hierarchy or by other tissue components. In bone, Gupta et al.[37] reported cooperative but unequal strain partitioning between tissue, collagen fibrils, and mineral, with an approximate ratio of 1:0.42:0.17. This corresponds to substantially less strain attenuation at the mineral level than observed here, consistent with differences in tissue architecture and in the structural level probed. In particular, both tendon and cartilage present with a much higher hydration as well as a larger presence of non-collagenous proteins such as proteoglycans, which both would strongly contribute to the dissipation of strain[17]. Strain partitioning in tendon has likewise been well documented by earlier synchrotron studies from Fratzl and co-authors[57,58,76], which showed that fibril strain remains well below tissue strain and described tendon as a hierarchical viscoelastic system in which fibrils and interfibrillar matrix act in a coupled manner. More recently, Sharma et al.[47] reported in rat Achilles tendon that fibrils experienced at most ~7% and collagen molecules ~2% of the applied tissue strain, in good agreement with the present data. Bianchi et al.[40] further showed that mechanical damage can differentially alter the response across hierarchical levels. Together, these studies support the view that tensile deformation is not transmitted directly across the hierarchy but is progressively attenuated. The findings of these studies are well in line with the results of the current study, further supporting the notion that the enthesis is a viscoelastic system in which the different hierarchical load-bearing components of different regions all act in a coupled matter. Finally, the particularly low strain observed in the HA crystals supports the interpretation that the mineral phase acts as a stiff reinforcing component that restrains deformation of the surrounding collagen matrix. As mineral reinforcement decreases from bone through MFc toward the interface, greater local deformation is therefore permitted. This coupling between regional specialization and hierarchical strain attenuation is summarized conceptually in Figure 5B.

In the enthesis, this attenuation is particularly relevant because load is not transferred within a single tissue, but across a graded tendon-to-bone interface with strong regional differences in mineralization, fibrocartilage composition, and local architecture[3,4,14–16,63,64]. The present study extends this view by showing that such strain partitioning is therefore not only hierarchical, but also spatially heterogeneous across the insertion itself. This is clearly observed by the gradually increased strain at preload as well as a more immediate and initially stronger deformation response towards the interface of all investigated structures.

This hierarchy-specific decoupling is consistent with deformation mechanisms acting across several structural levels. At the microscale, interfascicular and interfiber sliding can absorb substantial deformation before it is transferred to fibrils[77–79]. Within fibers, interfibrillar shear and rearrangement between neighboring fibrils can further reduce the strain reaching the fibrillar periodicity[54,79]. Within the fibril itself, limited sliding between staggered collagen molecules and gap region extension contributes to its elongation[39,56]. While information on molecular scale mechanics is sparse, extension of the collagen molecule has been hypothesized to arise from a combination of stretching and removal of molecular kinks[57]. In addition to these structural changes across the hierarchy, the non-collagenous matrix has been shown to play an important role[17,79]. Puxkandl et al.[76] proposed that tendon mechanics should be understood as the coupled response of fibrils and interfibrillar matrix, and Masic et al.[80] later showed that tendon collagen can carry osmotic pressure-induced tensile forces, indicating that fibrils exist in a hydrated matrix capable of generating internal prestress. This provides a plausible framework for how local changes in a proteoglycan-rich matrix could influence fibrillar strain state and load transfer through hydration and osmotic effects, rather than solely through direct tensile bridging between neighboring fibrils. Together, these findings underline the importance of the non-collagenous matrix in modulating load transfer by regulating fibril organization, hydration, sliding, and viscoelastic coupling.

### 3.3 Experimental context and limitations

At the tissue scale, the force range reached here is broadly consistent with previous mechanical studies on tendon entheses[3,6,22]. This supports the view that the present experiments probe a mechanically relevant part of the load-displacement curve. Together with the absence of major differences between the 1 and 15 min relaxation groups, this justifies pooling both groups for the structural analysis.

Radiation damage is an important constraint for synchrotron studies of hydrated biological tissues. Under the present conditions, each scan point received a total dose of 5.78 kGy across the four repeated maps, which was well below the earliest detectable changes in collagen scattering parameters observed in the repeated-exposure test (~289 kGy; SI Fig. S3). Radiation damage is therefore unlikely to have affected the present 2D measurements. However, it remains a central limitation for future 3D or time-resolved scattering studies, where the number of exposures per voxel would be substantially higher. The dose margin observed here nevertheless suggests that high-energy, low-exposure X-ray scattering may provide a viable route toward extending this approach to three dimensions.

Lastly, the findings of this study are based on 2D projections through a geometrically complex 3D interface. Because the exact size and borders of the anatomical regions vary in both coronal and sagittal directions, precise anatomical assignment remains limited and the measured strains cannot be converted directly into local stresses. Future studies should therefore combine in situ mechanics with 3D imaging. Recent synchrotron phase-contrast studies have already resolved microscale strains in the mouse enthesis[62], Achilles tendons[81], and cartilage and menisci[82], highlighting the potential of such approaches for extending the present framework. Such 3D extensions would, however, make radiation damage even more critical, because each voxel would receive substantially more exposure than in the present 2D mapping experiment. Thus, radiation damage remains a central constraint for future 3D scattering studies and scavenging strategies[83] to enhance the radiation budget need to be further investigated.

## 4 CONCLUSIONS

This study provides the first spatially resolved multiscale picture of deformation across the mouse Achilles tendon enthesis. Simultaneous mapping of the strain in collagen fibrils, collagen molecules, and hydroxyapatite mineral particles and crystals during *in situ* loading reveals structural gradients already at preload, with an earlier and stronger response of interface-adjacent tissues, and strong attenuation of strain across hierarchical levels. These results show that load transfer across the enthesis is both spatially heterogeneous and hierarchy-dependent, which support the view that this graded attachment accommodates deformation through region-specific load sharing and hierarchical strain partitioning. This attenuation of strain through the hierarchical structure further underlines the importance of the non-collagenous matrix such as proteoglycans in modulating load transfer. In doing so, the enthesis can mitigate stress concentrations and maintain mechanical integrity across the tendon-to-bone transition. More broadly, this framework provides a basis for future studies of three-dimensional and time-resolved interface mechanics, which may help guide the design of improved strategies for enthesis repair and bioinspired graded attachments.

## 5 MATERIALS AND METHODS

### 5.1 Samples

Eleven male C57BL/6J mice (12 weeks old) were used in this study; the right hindlimb from ten mice were used for mechanical testing, one was used for testing the setup and radiation damage. No animal experimental procedure was performed before euthanasia; therefore, no project authorization for an animal experiment was required for the present study as confirmed by the animal-welfare body of Aix-Marseille University. The animals were treated in accordance with the guidelines of ARRIVE and the French National Research Council for the Care and Use of Laboratory animals. The mice were housed in cages of up to five, with a light-dark cycle of 12 h and in temperature-controlled rooms (22°). A standard food diet was provided ad libitum. The mice were sacrificed after 2 weeks of acclimatization. Their right hindlimb was dissected to conserve the distal part of the tibia, the whole paw and the Achilles tendon together with the triceps surae unit (calf muscle). The plantaris tendon was removed and the samples were then stored frozen at -80°C until measurements.

### 5.2 *In situ* loading device

A loading device was designed and custom built (Fig 1.A). Displacement was enacted using a N310-11 piezo linear actuator (10µm step size) coupled to an E-861 NEXACT controller (Physik Instrumente (PI) GmbH & Co. KG, Germany). The displacement was measured by a RLB magnetic encoder head (1 µm resolution, 1µs edge separation) coupled with a magnetic scale. The displacement was encoded using a RLACC005 connector adapter and an E201 USB interface (Rotary and Linear Motion Sensors (RLS), Slovenia). The axial force was measured using a GSMTM-20N (PM Instrumentation, France) load cell and a GSV-2TSD-DI amplifier (ME-Meßsysteme GmbH, Germany). The bottom support was mounted on top of two perpendicular SLC-1730 linear piezo stages (SmarAct GmbH, Germany) to enable sample alignment. Grips were designed to sufficiently clamp the mouse Achilles tendon enthesis on both ends, similar to previously used protocols in the lab[3], and to minimize interference with the X-ray beam. A tube to maintain humidity was created by 3D printing a support and lid in PLA, the walls made of a 3525 Ultralene film (4µm, SPEX SamplePrep, USA) and fitted with an o-ring at its base. All devices were interfaced using a custom-written control software in Python using TANGO device servers, which were interfaced with the beamline control system Bliss.

**5.3 X-ray scattering**

Scanning SAXS and WAXS measurements were performed at the ID15A beamline at the European Synchrotron Radiation Facility (ESRF), France. A photon energy of 41 keV was selected using a multilayer monochromator with 0.37% energy bandwidth. A set of Al compound refractive lenses were used to focus the X-ray beam to a size of 20 x 25 µm, resulting in a photon flux of $10^{13}$ ph/s at the sample position. A two-pinhole collimator was used to reduce scattering background around the beam and a 3D printed flight tube, flushed with He was used to reduce the background (SI Fig S4.A). A 1 mm diameter beamstop made of gold was glued to the exit window of the flight tube.

Diffraction patterns were collected using a Dectris Pilatus 3 CdTe 2M detector at sample-detector distance of 726.5 mm, providing a q-range of 0.01 – 37 $nm^{-1}$. The setup was calibrated using a $CeO_2$ (NIST 674b) calibrant using the pyFAI python package[84].

After thawing, the samples were mounted by securing the foot on a support machined from polyetheretherketone (PEEK) using silk suture thread (2-0, BioSebLab, France) (SI Fig S4.B). The support was designed to position the foot at an angle of 135° relative to the Achilles tendon, mimicking the physiological loading axis of the murine Achilles tendon[3]. Using surgical scissors, the tibia was cut as close to the ankle joint as possible, and the foot was cut in order to not protrude outside of the diameter of the support. The tendon was clamped as close to the enthesis as possible, using sandpaper to avoid slipping. The mounted sample and grips were enclosed in a thin film tube sealed to the bottom support with an O-ring to avoid leakage. The sample was hydrated with phosphate buffered saline (PBS) and pieces of cotton wetted with PBS were placed at the bottom of the tube to maintain humidity during the measurements (SI Fig S4.B). The clamps with the tube were mounted in the device and the tendon aligned to the calcaneal bone in the sagittal and coronal plane by the two perpendicular linear piezo stages using a microscope (Dinion Color LTC0440 Camera, Bosch, US) fitted with an objective (Edmund Optics, UK) and a rotational stage (DRTM 90-D53-HiSM, OWIS GmbH, Germany).

Samples were preloaded to 2N by applying tensile displacement to the upper grip. Three incremental displacement steps of 200 µm were applied using a displacement rate of 0.17±0.03 mm/s followed by 1- (N = 5) or 15 minutes (N = 5) relaxation. At preload and after relaxation at each step, 2D SAXS/WAXS maps were acquired (adding ~1 minute/step).

2D SAXS/WAXS maps of the samples were acquired over a FOV of 1 x 1.5 $mm^2$ with a step size of 20 µm in both directions and an exposure time of 4 ms plus 1 ms readout time.

### 5.4 Data analysis

The analysis of the mechanical and scattering data was performed using custom-written Python codes. Similar to previous approaches[43,44,47], force was determined at the peak of each step from the tissue mechanical data,. Force relaxation ratio was determined as the ratio of the relative decrease in force from the peak compared to the end of the relaxation period. Fast relaxation time was estimated as the time from the peak until 60% of the relative relaxation in force had been reached. Stiffness was determined as the slope of a linear fit of the force-displacement data reaching each step. Additionally, the displacement was corrected using the nominal compliance of the load cell (5 µm/N).

The applied tissue strain was calculated as engineering strain from the displacement $d$ of the upper grip, corrected for load cell compliance and divided by the initial gauge length of the mounted sample at preload ($d_0$):

$$\varepsilon = \frac{d - d_0}{d_0} x100$$

The initial gauge length was estimated from microscope images as the distance between the upper clamp and the bottom support. This value represents the global applied strain of the whole tested specimen, rather than a local tissue strain. In the same manner, scattering-derived strains for a given step were calculated independently for each sample and region as the relative change in the estimated spacing compared with the corresponding value at preload.

The scattering patterns were masked to remove the beam stop and gaps between detector panels, along with hot and dead detector pixels. The scattering data were integrated using the PyFAI package[84] into 500 azimuthal bins between -180° and 180° to obtain $I(\chi)$ curves, as well as 2000 linearly spaced q-bins within 0.01-37 $nm^{-1}$ to obtain $I(q)$ curves. $I(\chi)$ curves were extracted within the q-range related to scattering of the collagen fibrils in the axial (0.28 ± 0.055 $nm^{-1}$) and lateral (4.35 ± 0.185 $nm^{-1}$) direction, the mineral particles (0.55 ± 0.055 $nm^{-1}$), the amino acids of the collagen molecules in the axial direction (21.75 ± 0.74 $nm^{-1}$), the HA (002) reflection (18.2 ± 0.185 $nm^{-1}$), i.e. along the c-axis (Fig 1.C). The peaks in the $I(\chi)$ curves were fitted with a Gaussian with linear background, the orientation was determined from the position of the fit and the degree of orientation estimated as the ratio between the anisotropic scattering and the total scattering (SI Fig S5.A)[85]. For further structural analysis, $I(q)$ curves were averaged in both directions of the determined orientation ± 15° (SI Fig S5.B), except for the mineral particle thickness analysis which utilized the full azimuthal range. Due to a heavy influence of the direct beam on the orientation analysis at low q, the amino acid orientation was used to retrieve the $I(q)$ curves for the collagen fibril axial parameters. Due to the strong influence in the $q$-region of the HA (310) from the bottom support in PEEK, this peak analysis was excluded from the results.

All $I(q)$ curves were background corrected with air scattering from pixels within the same scan and their peaks fitted with a Gaussian with linear background for d- or lattice spacing (peak position), heterogeneity (peak full-width-at-half-maximum, FWHM) and amount / intrafibrillar order (peak area), following previously described approaches[43,47]. Due to the few pixels around the beam stop contributing to the q-range of the 3rd order axial collagen fibril reflection, the 5th order reflection was instead used for axial d-spacing $d_{fibril\text{-}A}$, fitted in the q-range of 0.35-0.57 $nm^{-1}$. Lateral molecular packing within the fibrils $d_{fibril\text{-}L}$ was determined from the peak fit in the q-range of 2-7 $nm^{-1}$, the amino acid spacing of the collagen molecules $d_{amino}$ from 18-25 $nm^{-1}$ and the HA lattice spacing along the c-axis $d_{002}$ from 18-25 $nm^{-1}$. To estimate the thickness of the mineral particles $T$, a two-phase system with 50% mineral fraction[27,38] and predominantly platelet-shaped particles were assumed. As previously described[30,86], the mineral particle size was estimated from the Porod constant $P$ and invariant $J$ using:

$$T = \frac{4J}{\pi P} = \frac{4}{\pi P}\int_0^{\infty} q^2 I(q)\, dq$$

A linear fit of the q-range 1.1-2.0 nm$^{-1}$ in the Porod plot, i.e. $I(q^4)$, was extrapolated to $q$=0 to determine the Porod constant.

Regional and profile analysis were conducted on all estimated parameters. Prior to analysis, the collagen and mineral tissues were selected by using a threshold above the background based on 10% of the maximum peak area of each parameter and potential peak misfits removed by unreasonable d- or lattice-spacings. The maps of the collagen parameters were selected manually for the region closest to the mineral interface / unmineralized fibrocartilage (UFc) as well as the main part of the tendon (T) and the maps of the mineral parameters were masked manually for mineralized fibrocartilage (MFc), periosteal fibrocartilage (PFc), upper (UB) and lower (LB) halves of the calcaneal bone in the field of view. The masks were based on the approximate size and placement of the regions observed in histological[3,4] and inhouse µCT images. Average values per displacement step were generated by first averaging the values within each mask of each sample, then averaging all samples for each parameter at a given step.

The fibril contraction ratio for the collagen fibril was estimated as the ratio of lateral ($\varepsilon_{lateral}$) to axial ($\varepsilon_{axial}$) strain obtained from each displacement step:

$$fibril\ contraction\ ratio = -\frac{\varepsilon_{lateral}}{\varepsilon_{axial}}$$

The interface which separates the mineralized from the unmineralized tissues was identified as the overlap between MFc and UFc masks after binary dilation of 1 pixel. Distance maps from the interface were generated by combining the UFc and T masks for the collagen parameters and the MFc and LB masks for the mineral parameters. For each parameter of each scan, the values of all pixels with the same distance were averaged, resulting in an average profile as a function of distance to the interface. All samples were then averaged to generate a general distance profile for each displacement step.

**5.5 Radiation damage test**

A limiting factor during synchrotron experiments of biological samples is the radiation dose. A too high radiation dose will damage the tissues and strongly affect their mechanical properties. To make sure that our experimental setup was within the limits of radiation damage and to define them at the energy used in this experiment as well as for the previously unreported molecular scale, we conducted 10 000 repeated exposures of 4 ms plus 1 ms readout time in a point of the tendon with strong signal from one of the samples. The dose deposited on the sample was estimated as[87]:

$$dose = \frac{\mu_{en} N_0 \epsilon}{\rho} \cdot \tau$$

where $\mu_{en}/\rho = 0.07\ cm^2/g$ is the linear mass-energy absorption coefficient for soft tissue at 41.5 keV[88], $N_0 = 3.18 \cdot 10^{19}\ photons/s \cdot m^2$ is the incident photon density per scan spot (i.e. photon flux over the beam area), $\epsilon = 6.49 \cdot 10^{-15}\ J$ is the photon energy and $\tau = 5 \cdot 10^{-3}\ s$ is the exposure time (including 1ms detector readout time) per scan spot. This experiment thus resulted in a dose rate of 289 kGy/s. Therefore, the dose for each scan point through the full sample thickness was 1.44 kGy, resulting in a total dose of 5.78 kGy in each point from the 4 scans. Substantial changes in the fitted scattering parameters were not observed during the repeated exposure test for the axial d-spacing, lateral molecular packing and amino acid spacing until after approximately 3000, 200 and 2000 exposures, respectively. This corresponds to approximately 12, 1 and 8 s of total exposure, and 4.34, 0.29 and 2.89 MGy, respectively (SI Fig S3).

### 5.6 Statistical testing

Linear mixed effects analysis was used to test for statistically significant ($p < 0.05$) changes with displacement steps within a region (one-way) and the combined effect of both region and steps (two-way). In both cases a random effects model was used as a starting point. If the covariance of the fitted model was too low (< 0.01), no random effects were found in the data and thus a fixed effects model was used instead. The residuals of the fitted model were tested for normal distribution with the Shapiro-Wilks test and if not normally distributed, a robust model was used. When a significant effect was found, post-hoc tests were conducted at each displacement step between regions using the t-test if the data was normally distributed, otherwise the Mann-Whitney U-test.

## 6 ACKNOWLEDGEMENTS

We acknowledge the European Synchrotron Radiation Facility, Grenoble, France, for providing beamtime at the ID15A beamline under proposal LS3372, as well as their beamline staff for their excellent help before, during and after the beamtime. We acknowledge the Partnership for Soft Condensed Matter (PSCM) for support during the preparation of the experiment. M. M. gratefully acknowledges funding by the Austrian Science Fund (FWF) 10.55776/ESP2934524.

This project has received funding from the European Research Council (ERC) under the European Union's Horizon Europe research and innovation program (ERC StG "TexTOM", grant agreement No 101041871). Views and opinions expressed are those of the authors only and do not necessarily reflect those of the European Union or the European Research Council. Neither the European Union nor the granting authority can be held responsible for them.

The X-ray scattering datasets are available under the following DOI:

https://doi.esrf.fr/10.15151/ESRF-DC-2439293166

## SUPPORTING INFORMATION

# Structural gradients and strain partitioning across the mouse Achilles tendon enthesis revealed by *in situ* X-ray scattering

**Authors:** Isabella Silva Barreto[1], Moritz L. Stammer[1], Moritz P.K. Frewein[1], Claire Camy[2,3], Juraj Todt[4,5], Michael Meindlhumer[4], Jozef Keckes[4], Stefano Checchia[6], Sandrine Roffino[2], Martine Pithioux[2,3,7,†], Tilman A. Grünewald[1,†,*]

**Affiliations:**
[1] Aix-Marseille Université, CNRS, Centrale Med, Institut Fresnel, Marseille, France
[2] Aix-Marseille Université, CNRS, ISM, Institut des Sciences du Movement, Marseille, France
[3] Aix-Marseille Université, APHM, CNRS, ISM, Mecabio Facility, Anatomy Laboratory, Timone, Marseille, France
[4] Department of Materials Science, Montanuniversität Leoben, Franz Josef Straße 18, 8700 Leoben, Austria
[5] Erich Schmid Institute for Materials science, Austrian Academy of the Sciences, Jahnstraße 12, 8700 Leoben, Austria
[6] European Synchrotron Radiation Facility, Grenoble, France
[7] Aix-Marseille Université, APHM, CNRS, ISM, Sainte-Marguerite Hospital, Institute for Locomotion, Marseille, France

[†]These authors contributed equally

*To whom correspondence should be addressed:
Tilman A. Grünewald, Email: tilman.grunewald@fresnel.fr
Institut Fresnel
52 Avenue Escadrille Normandie Niemen, 13013 Marseille, France

## SUPPORTING RESULTS

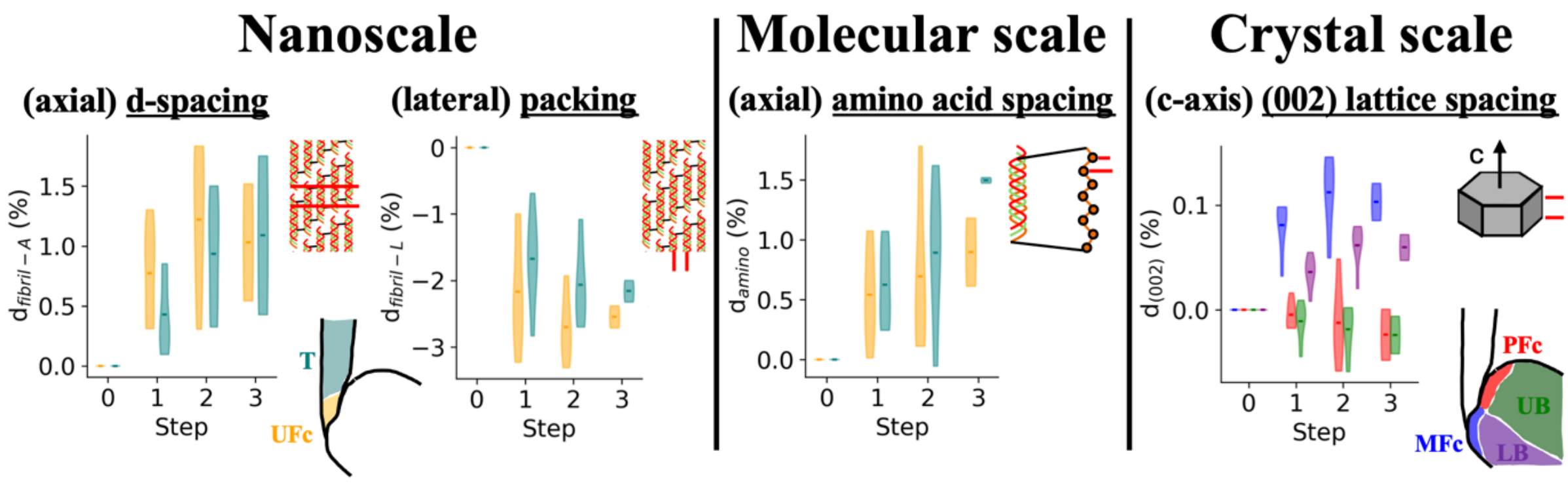

**Figure S1. Hierarchical strain partitioning.** Comparison between the strains observed at the nano-, molecular- and crystal scale within each region across displacement steps (0 = preload).

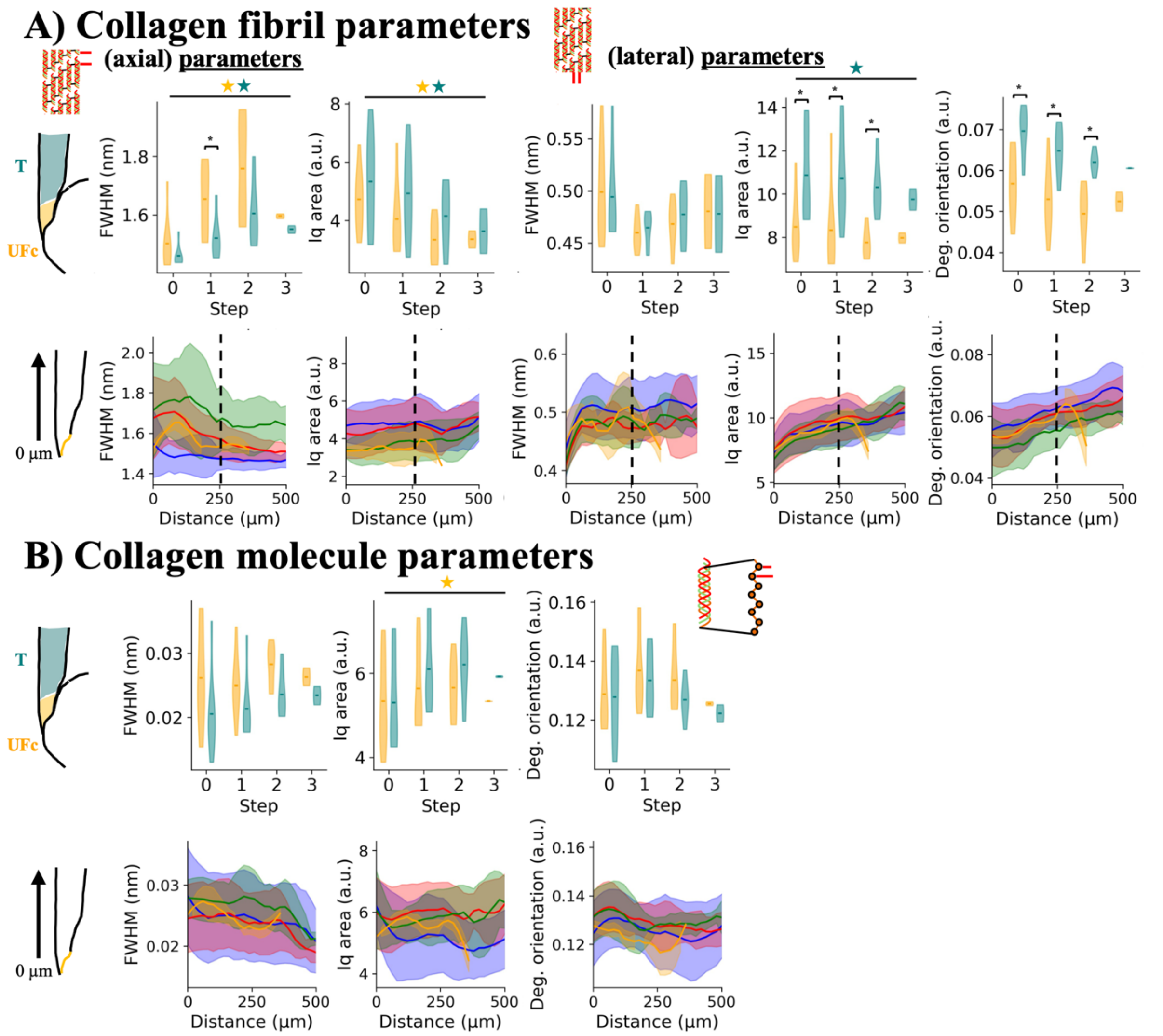

**Figure S2. Additional unmineralized tissue results.** A) Nanoscale collagen fibril axial and lateral and B) molecular scale collagen amino acid mechanical results within each region across displacement steps (0 = preload). Strain heterogeneity is indicated by FWHM and the I(q) area is related to the detected amount / order within the probed volume. Statistical significance ($p < 0.05$) is indicated with steps within a region (color-coded stars) and between regions at each step (asterisk).

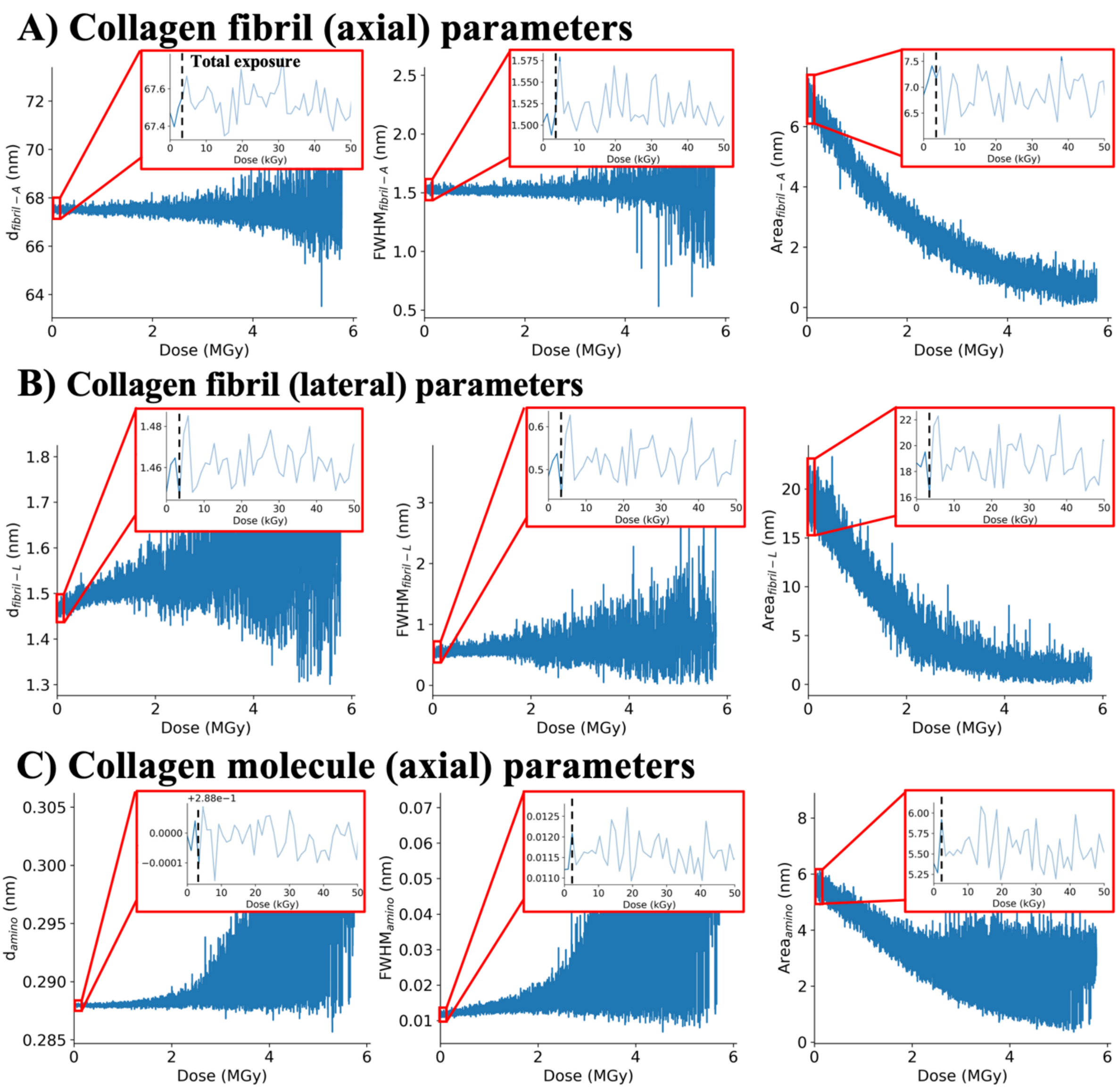

**Figure S3. Radiation test results.** Evolution of A) axial fibril, B) lateral fibril and C) axial amino acid spacing, heterogeneity (peak FWHM) and amount / order (peak area) with repeated exposures in the same spot of the Achilles tendon of 4ms exposure plus 1 ms readout time each. The evolution during the first 50 kGy is highlighted in the red boxes, with a dashed line indicating the total exposure in each point of each sample for the 4 maps acquired (5.78 kGy).

## SUPPORTING METHODS

**A) Experimental setup**

X-rays

Flight tube

Rotation stage

**B) Sample mounting**

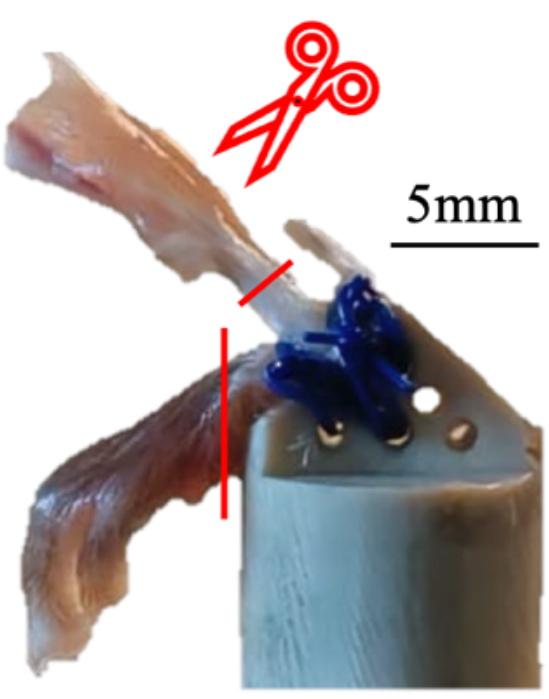

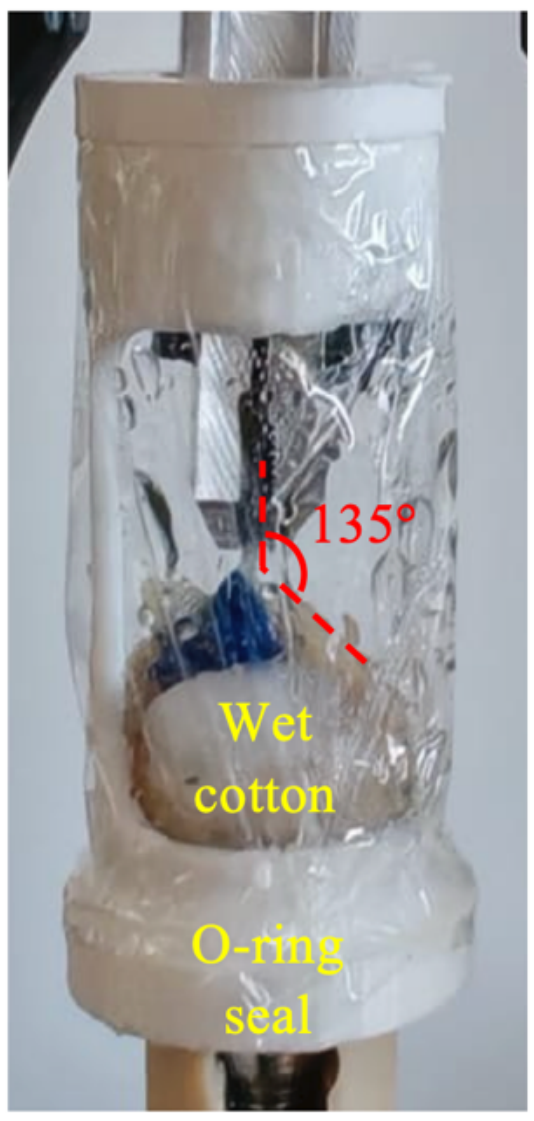

**Figure S4.** ***In situ*** **experimental setup.** A) Loading device mounted at the ID15A beamline, ESRF, France. B) Sample mounted on the bottom support and clamped inside the thin film tube.

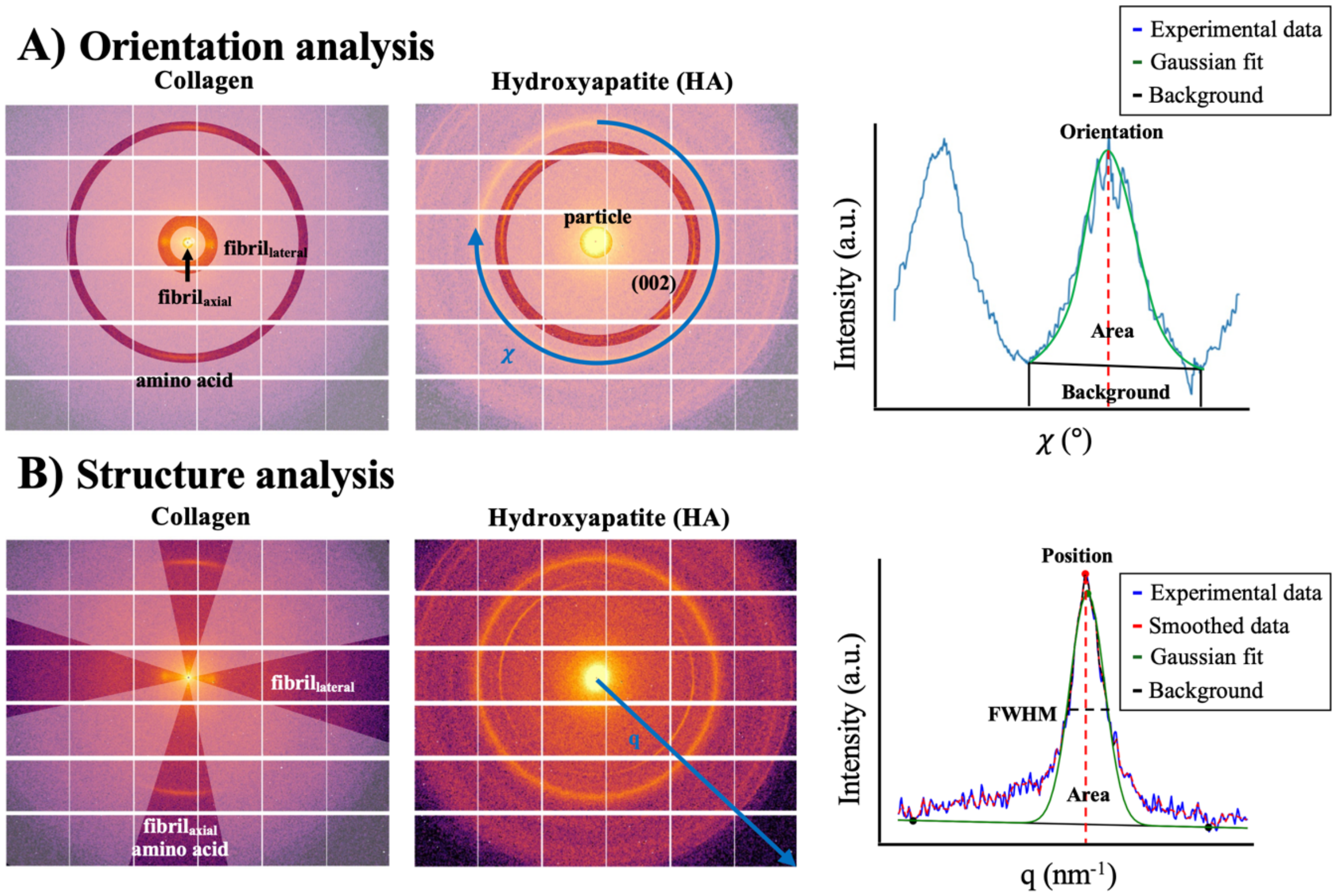

**Figure S5. Scattering data analysis.** A) Orientation analysis was performed for each parameter by integrating the scattering patterns azimuthally in the q-range around their respective peaks (non-shaded areas), creating $I(\chi)$ curves. The peaks were fitted with Gaussians using a linear background and the predominant orientation estimated from the fitted peak position. Degree of orientation was estimated from the ratio of the area under the fitted peak (anisotropic scattering) compared to the area of the total scattering (anisotropic scattering + background). B) Structural analysis was performed for each parameter by integrating the scattering patterns radially in both directions of the determined orientation ± 15° creating $I(q)$ curves, except for in the case of mineral particle thickness where it was averaged over the full azimuthal range. The $I(q)$ curves were background corrected with air scattering and the peaks fitted with Gaussians using a linear background. For each parameter except the particle thickness, the d- or lattice spacing was estimated from the fitted peak position, their heterogeneity from the width of the peak at half maximum (FWHM) and their amount from the peak area.